# High-Voltage Honeycomb Layered Oxide Positive Electrodes for Rechargeable Sodium Batteries

Chih-Yao Chen,[a] Josef Rizell,[b] Godwill Mbiti Kanyolo,[c] Titus Masese,[*,a,d] Yasmine Sassa,[b] Martin Månsson,[e] Keigo Kubota,[a,d] Kazuhiko Matsumoto,[*a,f] Rika Hagiwara[a,f] and Qiang Xu[*a,d]

Correspondence: titus.masese@aist.go.jp

[a] AIST-Kyoto University Chemical Energy Materials Open Innovation Laboratory (ChEM-OIL), National Institute of Advanced Industrial Science and Technology (AIST), Sakyo-ku, Kyoto 606-8501, Japan

[b] Department of Physics, Chalmers University of Technology, SE-412 96 Göteborg, Sweden

[c] Department of Engineering Science, The University of Electro-Communications, 1-5-1 Chofugaoka, Chofu, Tokyo 182-8585, Japan

[d] Department of Energy and Environment, Research Institute of Electrochemical Energy (RIECEN), National Institute of Advanced Industrial Science and Technology (AIST), Ikeda, Osaka 563-8577, Japan

[e] Department of Applied Physics, School of Engineering Sciences, KTH Royal Institute of Technology, Roslagstullsbacken 21, SE-106 91 Stockholm, Sweden

[f] Graduate School of Energy Science, Kyoto University, Yoshida, Sakyo-ku, Kyoto 606-8501, Japan

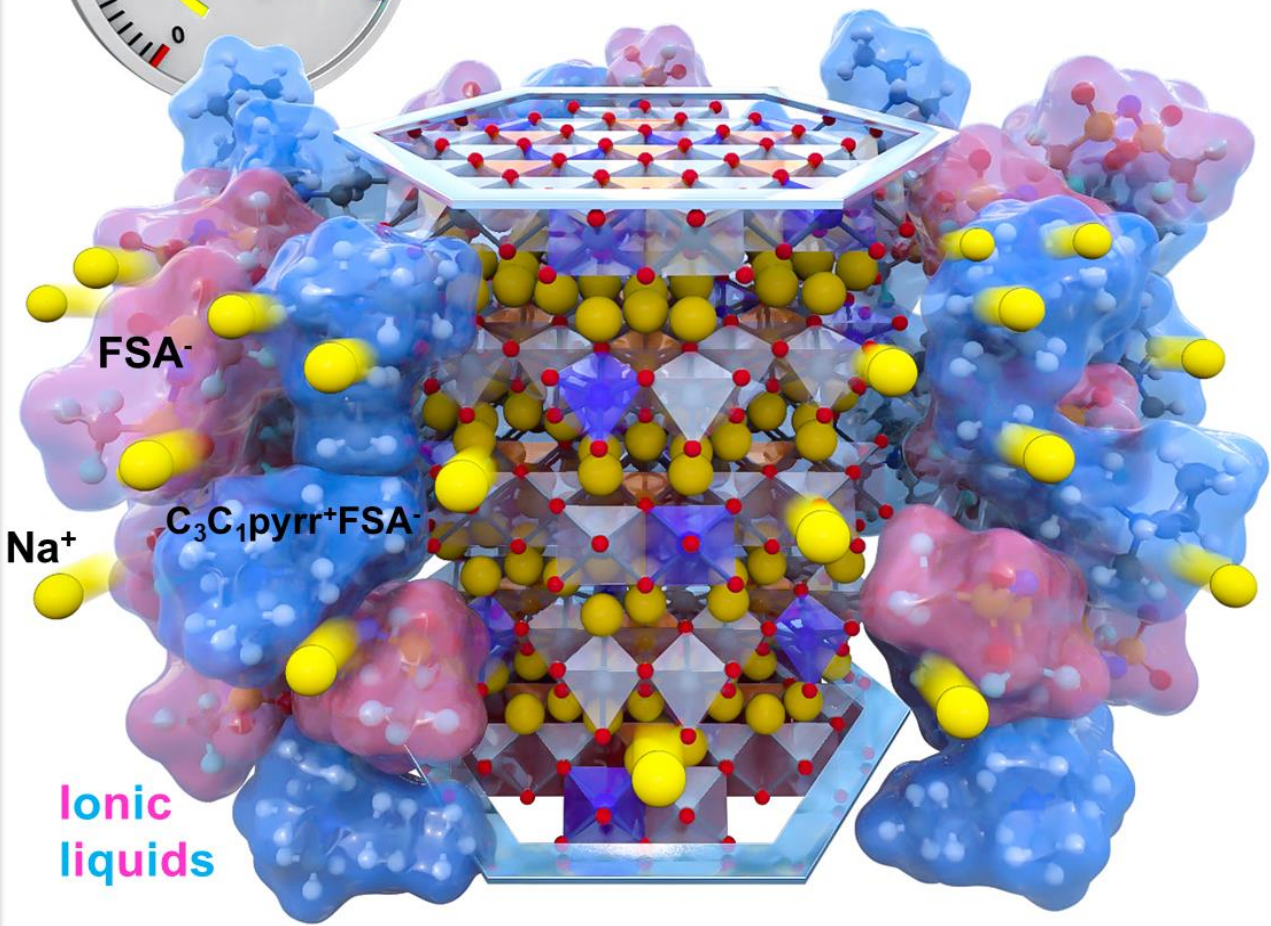

**Abstract**

Natural abundance, impressive chemical characteristics and economic feasibility have rekindled the appeal for rechargeable sodium (Na) batteries as a practical solution for the growing energy demand, environmental sustainability and energy independence. However, the scarcity of viable positive electrode materials remains a huge impediment to the actualization of this technology. In this paper (communication), we explore honeycomb layered oxides adopting the composition $Na_2Ni_{2-x}Co_xTeO_6$ ($x$ = 0, 0.25 and 0.50) as feasible positive electrode (cathode) materials for rechargeable sodium batteries at both room- and elevated temperatures using ionic liquids. Through standard galvanostatic assessments and analyses we demonstrate that substitution of nickel with cobalt in $Na_2Ni_2TeO_6$ leads to an increase in the discharge voltage to nearly 4 V (*versus* $Na^+$/Na) for the $Na_2Ni_{2-x}Co_xTeO_6$ family of honeycomb layered oxide materials, which surpasses the attained average voltages for most layered oxide positive electrode materials that facilitate Na-ion desertion. We also verify the increased kinetics within the $Na_2Ni_{2-x}Co_xTeO_6$ honeycomb layered oxides during operations at elevated temperatures which lead to an increase in reversible capacity of the rechargeable Na battery. This study underpins the doping of congener transition metal atoms to the honeycomb structure of $Na_2Ni_2TeO_6$ in addition to elevated-temperature operation as a judicious route to enhance the electrochemical performance of analogous layered oxides.

**INTRODUCTION**

As the contemporary world grapples with the effects of climate change and the approaching threshold of a future energy crisis, an imminent sense of prominence has been ushered into the world of rechargeable batteries. Owing to their high energy density and desirable charge cyclability, lithium-ion batteries have seen momentous growth in popularity as energy storage systems for large-scale use especially in this age of rapid industrial expansion and dynamic technological evolution. However, continued demand for economic sustainability, energy independence and environmental sensitivity have impaired development of lithium-ion batteries due to high costs and the scarcity of lithium mineral reserves.

Accordingly, rechargeable sodium (Na) batteries have shown great potential as future alternative capacious energy storage systems due to their low cost of development and the affluence of sodium mineral reserves. Furthermore, the redox potential of Na ( –2.71 V), being relatively lower than that of the standard hydrogen electrode (SHE), points towards prospects of a high energy density battery system if matched with high-voltage positive electrodes.[1] Although, previous studies have yielded an assortment of potential anode materials such as carbon-based materials, the advancement high performance rechargeable sodium batteries is heavily impeded by scarcity of high energy density positive electrode materials. As such, tremendous efforts towards the identification and development of feasible positive electrode materials remain paramount.

Similarities in the operation mechanisms of both sodium and lithium batteries have allowed adoption of layered transition metal oxides with composition stoichiometry analogous to those of the current rechargeable lithium batteries. As such, owing to their facile syntheses, great structural diversity and good electrochemical performance, $Na_xMO_2$ (where $x \leq 1$; $M$ = transition metal(s)) layered oxides, have been earmarked as appealing positive electrode candidates.[2–6] However, even with the great $Na^+$ electrochemistry, a vast majority of the reported layered oxides tend to exhibit relatively low operation voltages

(typically below 3.0 V versus $Na^+/Na$), making them unsuitable for high voltage regimes.[7]

In pursuit of superior positive electrodes for high voltage performance, the rising class of honeycomb-layered oxides adopting the composition $Na_x(MM')O_2$ (where $M'$ = Te, Sb, Nb, Ru, Bi, *etc*.) has emerged. This family of oxides has been seen to thrive in high voltage operations compared to the classic $Na_xMO_2$ compositions.[8–12] Partial substitution of $M$ in $Na_xMO_2$ with high-valent $M'$ metals (with fully occupied $4d$ orbitals) yields a layered oxide structure with a honeycomb configuration of $M$ metal atoms around $M'$ atom; endowing the honeycomb layered oxide not only with enhanced thermal stability but also with high-voltage performance.[13]

In a concept known as the "inductive effect", when an appropriate content of high-valent $M'$ metals is introduced into positive electrodes consistent of $Na_x(MM')O_2$ honeycomb layered oxide, the $M$–O bonds become more ionized; hence more energy is required to oxidize $M$ metal cations during Na-ion desertion. As a result, honeycomb layered oxides such as $Na_2Ni_2TeO_6$ (or equivalently as $Na_{2/3}Ni_{2/3}Te_{1/3}O_2$) demonstrate preponderant voltages (around 3.6 V) in comparison with $Na_xMO_2$.[12] As seen in seminal works performed on related positive electrode materials for rechargeable lithium and potassium batteries, partial substitution or doping of $M$ in $Na_x(MM')O_2$ with congener transition metal atoms has become a pertinent strategy to further increase the voltage of honeycomb layered oxides and expand their diversity.[14–17]

Another approach to unlocking the potential of positive electrodes is cycling at higher temperatures. Previous research on lithium and potassium batteries has also shown that increasing the temperature of operations boosts the mobility of ions within the electrodes. However, some organic electrolytes decompose during operations due to high temperatures thus utilization of electrolytes (such as ionic liquids or all-solid-state compounds) stable at high voltages and high temperatures would be propitious for their operation not only at high-voltage regimes but also at elevated temperatures.[18–21]

Herein, we report a novel assessment of the electrochemical performance of honeycomb layered oxide positive electrodes adopting the composition $Na_2Ni_{2-x}Co_xTeO_6$ (where $x$ = 0.25 and 0.50) using ionic liquids at both room- and elevated temperatures. $Na_2Ni_{2-x}Co_xTeO_6$ demonstrates ability to facilitate reversible Na-ion desertion with a high voltage close to 4V (*versus* $Na^+$/Na) at room temperature. Subsequently, a significant improvement in the capacity is exhibited at elevated temperatures (50 and 75 °C), surmising that Na-ion kinetics in $Na_2Ni_{2-x}Co_xTeO_6$ can be remarkably increased at elevated temperatures. This is a notion that could give more headway for the advancement of rechargeable sodium batteries as alternative energy storage devices for high voltage and high temperature operations.

**RESULTS AND DISCUSSION**

Honeycomb layered oxides adopting the composition $Na_2Ni_{2-x}Co_xTeO_6$ ($x$ = 0, 0.25 and 0.50) were synthesized via the conventional solid-state ceramics route at 800 °C, as explicated in the **Supplementary Information**. **Figure 1** shows the X-ray diffraction (XRD) patterns of the as-synthesized $Na_2Ni_{2-x}Co_xTeO_6$ ($x$ = 0, 0.25 and 0.50), fully indexed in the $P6_3/mcm$ hexagonal space group using the Rietveld refinement method. The refined atomic parameters for the XRD patterns are shown in **Tables S1**, **S2** and **S3** (**Supplementary Information**). The $Na_2Ni_{2-x}Co_xTeO_6$ ($x$ = 0, 0.25 and 0.50) samples possess honeycomb layered frameworks with Na atoms sandwiched between honeycomb slabs of transition metal atoms (Ni and Co) and tellurium (Te) that are coordinated with oxygen, as shown in **Fig. 1** inset. The honeycomb slabs entail the arrangement of $TeO_6$ with $Ni(Co)O_6$ octahedra in a manner that Te atoms are surrounded by six Ni(Co) atoms (honeycomb configuration). Na atoms are coordinated with six oxygen atoms of the adjacent honeycomb slabs in a prismatic coordination and partially occupy three crystallographically independent sites (designated as Na1, Na2 and Na3).

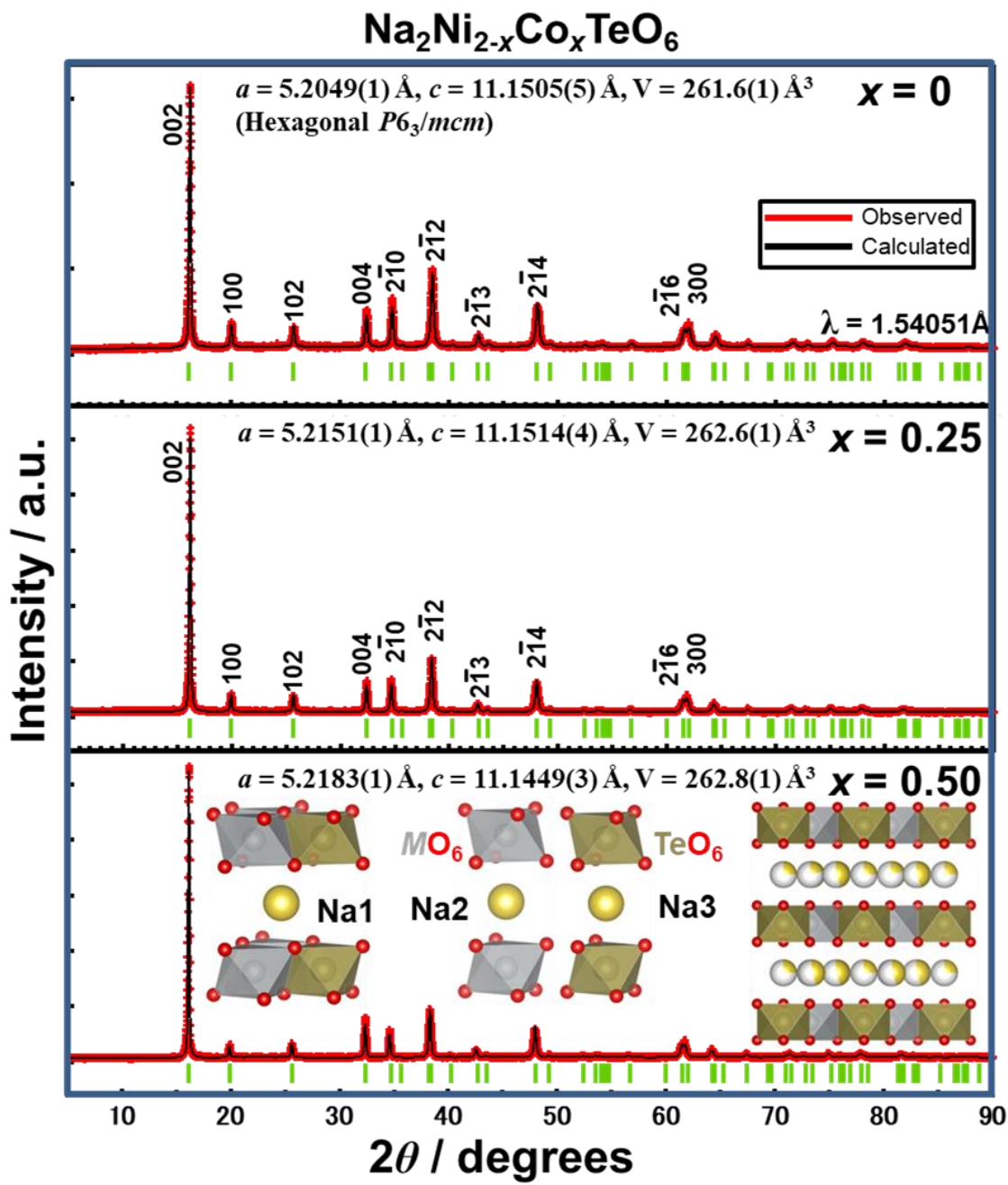

**Figure 1** Rietveld refinement results of conventional X-ray diffraction (XRD) data for Na$_2$Ni$_{2-x}$Co$_x$TeO$_6$ ($x$ = 0, 0.25, and 0.50), indexed in the hexagonal $P6_3/mcm$ space group. A polyhedral view of the honeycomb layered structure is embedded in the figure, in which Na is in yellow, (Ni, Co) in grey, Te in olive and O atoms in red. Na atoms occupy crystallographically distinct sites denoted as Na1, Na2 and Na3. Further details regarding the refinement protocols and attained reliability values are provided as **Supplementary Information.**

Increasing the amount of Co substitution Na$_2$Ni$_{2-x}$Co$_x$TeO$_6$ ($x$ = 0.25 and 0.50) does not significantly alter the occupancy of Na in the aforementioned respective sites (**Tables S1, S2** and **S3**). However, lattice parameters illustrate a propensity for increase in the size dimensions of transition metal slabs with an increase in the amount of Co substitution leading to an overall increase in the volume of the structure. This is clearly reflected by the *a*-axis parameter which represents the size dimensions of the transition metal honeycomb slabs. This is not unprecedented as Co$^{2+}$ in an octahedral coordination with oxygen (CoO$_6$) has a larger Shannon-Prewitt ionic radius than Ni$^{2+}$, thereby leading to an increase in the overall volume of the layered structure.[22] Moreover, partial substitution of Co in Na$_2$Ni$_2$TeO$_6$ has a profound effect on the synthesized powder color (see **Fig. S1**), demonstrating a possible application of cobalt substitution for optic purposes.[23] Scanning electron microscopy (SEM) images shown in **Fig. S2** reveal that the synthesized Na$_2$Ni$_{2-x}$Co$_x$TeO$_6$ compositions possess wide micrometric particle size distribution with irregular morphology. It is worthy to note that the samples show more defined crystalline facets and larger particle sizes upon Co-substitution as compared to the unsubstituted Na$_2$Ni$_2$TeO$_6$.

Electrochemical behavior of Na$_2$Ni$_{2-x}$Co$_x$TeO$_6$ as positive electrode materials was evaluated using Na half-cell counter electrodes. An (electro)chemically stable ionic liquid electrolyte comprising Na[FSA]–[C$_3$C$_1$pyrr][FSA] (FSA: bis(fluorosulfonyl)amide; C$_3$C$_1$pyrr: *N*-methyl-*N*-propylpyrrolidinium) with a molar ratio of 2:8 was adopted, as it demonstrates low charge-transfer, low interfacial resistances with a variety of electrode materials and stability in high temperature and high voltage regimes, allowing an accurate performance evaluation in a wide temperature range.[20,21]

In order to ascertain the correlation between the redox reaction and the temperature of operation, cyclic voltammetry (CV) of Na$_2$Ni$_{2-x}$Co$_x$TeO$_6$ ($x$ = 0, 0.25 and 0.50) was performed at a scanning rate of 0.1 mV s$^{-1}$ at 25 and 75 °C and the profiles are shown in **Fig. 2**. The samples exhibit redox peaks of above 3.2 V at both temperatures. However, sharper peaks and diminished peak separation are observed at 75 °C, suggesting that the redox reaction of $Na_2Ni_{2-x}Co_xTeO_6$ is thermally activated to a significant extent.[20,24] Another salient feature revealed in the CV profiles is that the number of oxidation peaks of $Na_2Ni_2TeO_6$ appear to decrease with the increase in operating temperature as well as with increasing amount of Co substitution. This phenomenon can be ascribed to phase transitions caused by the sliding of the honeycomb ($[Ni_{2/3}Te_{1/3}]O_2$) layers concomitant with the ordering of Na due to the vacancies created during Na-ion desertion, which would be suppressed by the increase of Co content and operation temperature.[8]

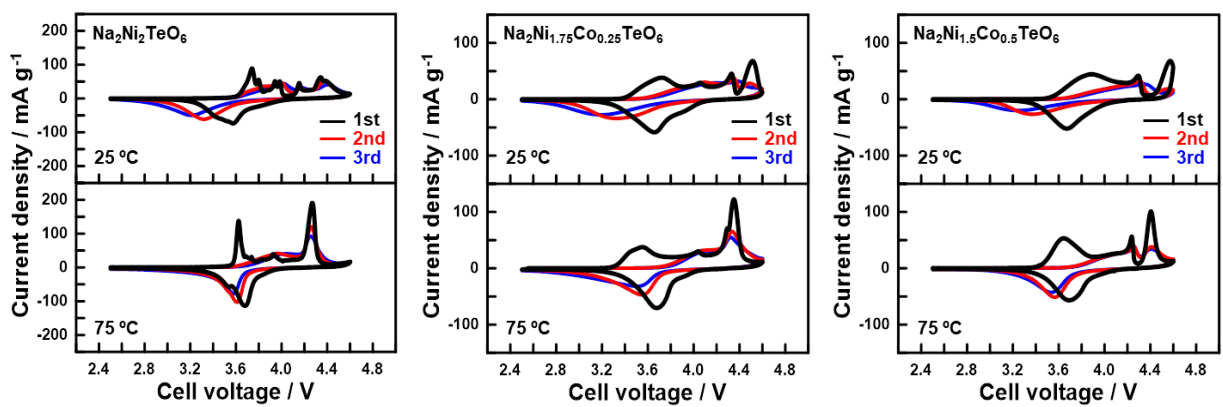

**Figure 2** Cyclic voltammograms of $Na_2Ni_{2-x}Co_xTeO_6$ ($x$ = 0, 0.25, and 0.50) electrodes at a scanning rate of 0.1 mV s$^{-1}$ in the voltage range of 2.5–4.6 V. The measurements were taken at 25 °C and 75 °C. A number of redox peaks reminiscent of phase transitions, which tend to be smeared and suppressed upon Co substitution, predominantly occur in the first cycle. Note that different scales have been used for each composition for clarity.

Typical voltage-capacity ((dis)charge) curves at a voltage range of 2.5–4.35 V measured at a current density equivalent to C/20 rate (6.925 mA g$^{-1}$) and at temperatures of 25, 50 and 75 °C are illustrated in **Fig. 3**. A theoretical capacity of approximately 138.5 mAh g$^{-1}$ was set as the standard for full Na removal from $Na_2Ni_{2-x}Co_xTeO_6$. In concordance with the CV result, unsubstituted $Na_2Ni_2TeO_6$ electrodes exhibit stepwise voltage profiles at 25 °C that correspond to multiple electrochemically-driven phase transitions.[13] Sloping voltage profiles are exhibited upon increasing the operation temperature to 50 and 75 °C, especially at the profile regimes corresponding to the removal of 1/4 units of Na per formula (commensurate to a capacity of 34.6 mAh g$^{-1}$). Higher configuration degrees of freedom in a thermalized system presumably perturb the ordering of Na in the structure, resulting in the gently sloping voltage profiles.[25] Sloping voltage profiles are generally viewed as more suitable as they allow better monitoring of the battery state-of-charge (SOC).[26] A reversible capacity of around 84–87 mAh g$^{-1}$ is attained over the entire temperature range, which is equivalent to ~1.26 Na desertion per formula unit (translating to *ca.* 53% utilization of the theoretical capacity of $Na_2Ni_2TeO_6$).

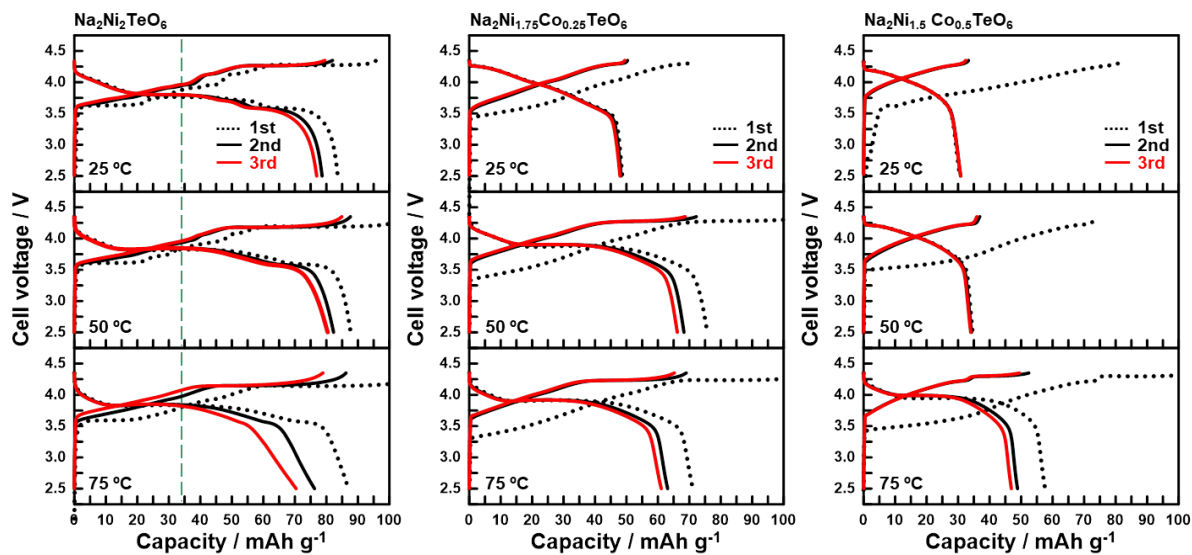

**Figure 3** Voltage-capacity plots of $Na_2Ni_{2-x}Co_xTeO_6$ (*x* = 0, 0.25, and 0.50) measured at a current density of 6.925 mA g$^{-1}$ (C/20) in the voltage range of 2.5–4.35 V at 25 °C, 50 °C and 75 °C. The first three cycles for each material composition are shown. The green dotted line indicates the capacity regime equivalent to the removal of 1/4 units of Na per formula (commensurate to a capacity of 34.6 mAh g$^{-1}$). The average voltage tends to increase upon Co substitution, albeit with a capacity decrease that can be compensated for by raising the operating temperature.

Upon partial substitution of Ni with Co in $Na_2Ni_2TeO_6$, the electrochemical behavior at both room- and elevated temperatures are notably different; the limited performance displayed at room temperature is significantly improved at elevated temperatures. Furthermore, a new voltage plateau appears at *ca.* 4.2 V at 50 °C. The capacity regime of this plateau increases with increasing temperature, leading to an increase in the reversible capacity from 49 mAh g$^{-1}$ (at 25 °C) to 71 mAh g$^{-1}$ (at 75 °C). For $Na_2Ni_{1.5}Co_{0.5}TeO_6$, the electrochemical performance is more prominently dependent on temperature, the reversible capacity increasing from 31 mAh g$^{-1}$ (at 25 °C) to 58 mAh g$^{-1}$ (at 75 °C) as shown in **Fig. 3**. Data relating to the capacity retention and coulombic efficiency are shown in **Fig. S3**. These results imply that the Na-ion desertion of these classes of materials is kinetically-restricted at ambient temperature.[26] The sluggish electrode kinetics observed at ambient temperatures (25°C) are in accord with deductions from previous electrochemical studies on honeycomb layered oxides.[12,13,16,17] The proportionally decreased reversible capacity with increasing Co incorporation, however, signifies that the electrochemical activity in this series of material solely arises from the Ni redox process. In fact, fully substituted $Na_2Co_2TeO_6$ hardly demonstrates Na-ion desertion.[12]

To preclude factors that may limit the redox process of Co, a lower current density (C/50 (2.77 mA g$^{-1}$)) and a higher upper cut-off voltage range (4.5 V) was applied for $Na_2Ni_{1.75}Co_{0.25}TeO_6$. As shown in **Fig. S4**, no increase in the capacity is observed, supporting our supposition pertaining to the inactive redox nature of Co. Notable is also that for all the three samples of $Na_2Ni_{2-x}Co_xTeO_6$ ($x$ = 0, 0.25 and 0.50), voltage profiles observed in the initial cycles are remarkably distinct from those of the subsequent cycles. This characteristic can be attributed to the structural rearrangements occurring during the initial cycle.[27] Trying to delve further into the structural rearrangement process calls for structural analyses based on synchrotron X-ray diffraction, which falls under the scope of future study.

In order to clarify the influence of Co substitution on the attained voltage plateau, differential voltage-capacity plots were acquired from the initial discharge curves and are shown in **Fig. 4** (corresponding curves are further provided in **Fig. S5**). Evidently, the differential voltage-capacity peak shifts to a higher voltage with increasing amount of Co substitution. The peaks of $Na_2Ni_{1.75}Co_{0.25}TeO_6$ and $Na_2Ni_{1.5}Co_{0.5}TeO_6$ shift to around 3.90 and 3.95 V, respectively, from that of $Na_2Ni_2TeO_6$ at 3.82 V. The average working voltages of the compounds of these compositions are calculated by dividing the area covered by the curve with the discharge capacity resulting in 3.71 and 3.74 V, respectively, which are record-high voltages attained for layered oxides operating on $Ni^{2+/3+}$ redox process (*viz.*, $NaNi_{0.5}Ti_{0.5}O_2$ (3.10 V),[28] $Na_3Ni_2BiO_6$ (3.10 V),[12] $Na_3Ni_2SbO_6$ (3.30 V),[10] $Na_2Ni_2TeO_6$ (3.55 V), $Na_{2/3}Ni_{1/3}Mn_{2/3}O_2$ (3.60 V),[2] and $Na_{0.7}Ni_{0.35}Sn_{0.65}O_2$ (3.70 V)[29]). The improvement in the capacity retention observed at elevated temperatures for Co-substituted $Na_2Ni_{2-x}Co_xTeO_6$ honeycomb layered oxide materials further demonstrates another beneficial aspect of Co in suppressing phase transformation (**Fig. S3**), as has also been previously observed in Co-substituted $Na_{0.67}MnO_2$[30] and $Na_{0.7}Mn_{0.7}Ni_{0.3}O_2$.[2,31]

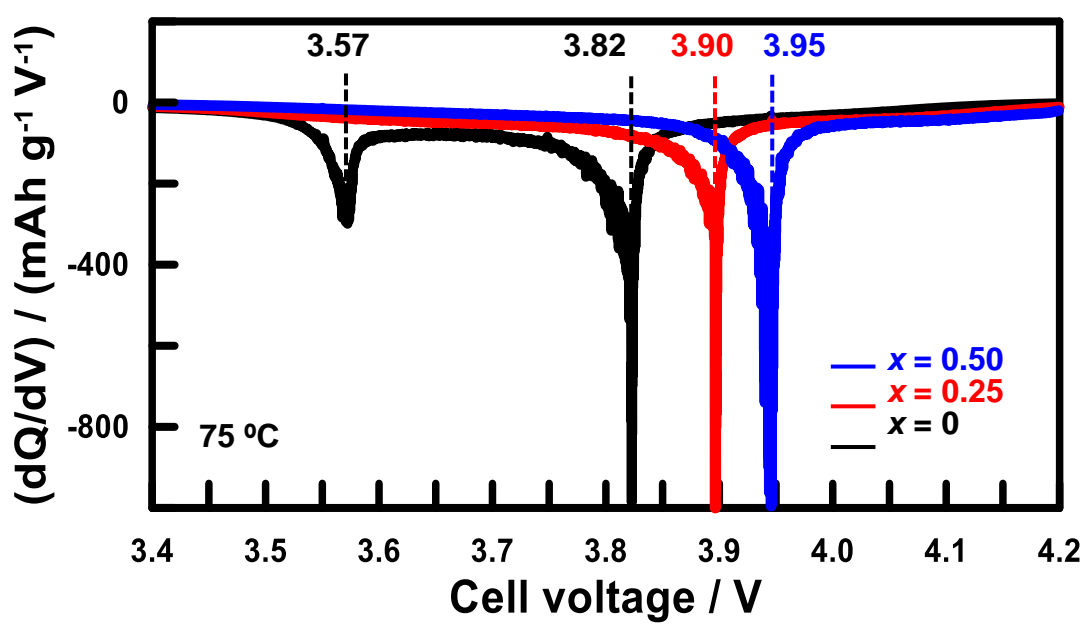

**Figure 4** Differential capacity ($\delta Q/\delta V$) plots of $Na_2Ni_{2-x}Co_xTeO_6$ ($x = 0$, 0.25, and 0.50) at 75 °C, indicating a systematic increase in the voltage with partial substitution of Co. These curves are derived from the first discharge voltage profiles shown in **Figure 3**.

At this juncture, it is imperative to further discuss the electrochemistry behind the effect of partial substitution of cobalt (Co) in $Na_2Ni_{2-x}Co_xTeO_6$. While Te does not participate in the redox process accompanying the desertion of alkali metal cations in related honeycomb layered oxides,[9,13,14,16] its presence in, for instance $Na_2Ni_2TeO_6$, has manifold effects of increasing the ionic character of Ni–O bond: Firstly, the high covalency of Te–O bonds withdraws more electrons and weakens the covalency of the neighboring Ni–O bonds (hence, becoming more ionic) as a result of the inductive effect.[32] Secondly, $Te^{6+}$ ($[Kr]4d^{10}$) has fully occupied $d$ electrons and therefore hardly interacts with O, leaving orbital hybridization primarily occurring between oxygen and transition metals that model the covalency of O–Ni bonds.[29] As a consequence, more energy is required to activate the Ni redox processes, culminating in an increase of the electrode voltages. Substitution of Ni in $Na_2Ni_2TeO_6$ with even a miniscule amount of Co (as demonstrated in the present study) could aid in the redistribution of the electronic density between O and Ni and further tailor the redox potential of Ni in $Na_2Ni_{2-x}Co_xTeO_6$ honeycomb layered oxide materials. Detailed theoretical computations regarding other compositions or other possible material substitutes are not covered in the current scope of work.

Honeycomb layered oxides such as $Na_2Ni_2TeO_6$ have been recognized to demonstrate fast Na-ion conduction, owing to their wide interlayer distances and two-dimensional Na-ion diffusion pathways.[33, 34] However, from the present study, it is evident that Na-ion desertion in $Na_2Ni_{2-x}Co_xTeO_6$ is rather a sluggish process that can be enhanced significantly by increasing the operation temperature. The discrepancy in performance presumably arises mainly from the exclusion of electrochemically-driven phase transformation during Na-ion transport, since additional energy barrier is required for phase boundary movement leading to non-negligible kinetic hindrance as observed.[24] Accordingly, it is demonstrably clear that cobalt substitution for nickel creates a new avenue for high performance elevated temperature battery systems.

Pending work remains to optimize the electrochemical performance of $Na_2Ni_{2-x}Co_xTeO_6$, strategies of which include *inter alia* manipulations of the stoichiometric amount of Na, particle size reduction (nanosizing) and carbon-coating. From a sustainability perspective, Na remains one of the most abundant metals available. Tellurium (Te) has long been known as a major by-product of copper refining and has been employed in megawatt-scale cadmium-tellurium (CdTe) solar photovoltaic arrays; therefore, Te precursors are readily obtainable. Also, there have been reports on non-terrigenous Te, but the mineral data remains scarce. As such, future work may also entail reduction or substitution of Te as a constituent element of rechargeable Na batteries due to prohibitive costs and poor environmental sustainability.

**CONCLUSION**

Recapitulating our central findings, honeycomb layered oxides adopting the stoichiometric formula $Na_2Ni_{2-x}Co_xTeO_6$ with variant amounts of substituted cobalt, were investigated as potential high-voltage positive electrode materials for rechargeable Na batteries. Taking full advantage of the chemically stable ionic liquids as the electrolytes, their electrochemical performance was substantially enhanced during operations at elevated temperatures. Particularly, $Na_2Ni_{1.75}Co_{0.25}TeO_6$ shows a reversible capacity of 71 mAh g$^{-1}$ with an average working voltage of 3.71 V (versus Na$^+$/Na) at 75 °C. This work demonstrates that a rational compositional design of $Na_2Ni_2TeO_6$ can be a simple yet effective route to uplift the redox voltage of Ni (or other transition metals) in honeycomb layered oxides.

## Acknowledgements

This work was conducted through the support of the National Institute of Advanced Industrial Science Technology (AIST) and Japan Prize Foundation. Y.S. and M.M. acknowledge support from the Swedish Research Council (VR) via a Starting Grant (Dnr. 2017-05078) and Neutron Project Grant (Dnr. 2016-06955), respectively, and J.R. is funded by Chalmers Area of Advance (Materials Science). We acknowledge that this paper was proofread and edited through the support provided by Edfluent services.

# Electronic Supplementary Information

## Experimental

*Synthesis and material characterization of $Na_2Ni_{2-x}Co_xTeO_6$*

Polycrystalline samples of $Na_2Ni_{2-x}Co_xTeO_6$ ($x$ = 0, 0.25 and 0.50) were synthesized using the high-temperature solid-state reaction. A stoichiometric mixture of $TeO_2$ (Aldrich, purity of ≥99.0%), NiO (High Purity Chemicals, purity of 99%), CoO (High Purity Chemicals, purity of 99%) and $Na_2CO_3$ (Rare Metallic, 99.9%) was thoroughly pulverised, then pelletised and subsequently calcined at 800 °C for 24 hours in air. Conventional XRD were measured using a D2 Phaser Bruker diffractometer at a Cu-$K\alpha$ wavelength, to confirm the purity of the samples. Structural refinements of $Na_2Ni_{2-x}Co_xTeO_6$ were performed using the Rietveld refinement protocol implemented in the JANA 2006 program and the crystal structure visualization was done using the VESTA software.[1,2] Morphologies of the powder product were analyzed using a scanning electron microscope (JSM-6510LA, JEOL).

*Electrochemical measurements*

$Na_2Ni_{2-x}Co_xTeO_6$ was mixed with acetylene black (AB) and polyvinylidene fluoride (PVDF) in a weight ratio of 70:25:5. A uniform slurry was made by suspending the mixture in *N*-methyl-2-pyrrolidinone, which was then coated onto aluminum foil with a mass loading of 4–6 mg cm$^{-2}$. The composite electrodes vacuum-dried at 120 °C overnight were used as the working electrode. Metallic sodium discs (Sigma-Aldrich, purity 99.95%) pressed onto aluminum current collectors were used as the counter electrodes. An ionic liquid Na[FSA]–[C$_3$C$_1$pyrr][FSA] (FSA: bis(fluorosulfonyl)amide; C$_3$C$_1$pyrr: *N*-methyl-*N*-propylpyrrolidinium) with a molar ratio of 2:8 (0.983 M at 25 °C) was used as the electrolyte due to its stable electrochemical properties over a wide temperature range.[3–5] Glass microfiber filters (GF/A, Whatman) vacuum impregnated with the electrolyte were used as the separator. All electrochemical measurements were performed using 2032-type coin cells assembled in an Ar-filled glove box with a Bio-Logic SP300 potentiostat or a Hokuto Denko HJ1001SD8 charge–discharge unit. The current density was converted into a $C$-rate based on the theoretical capacity of 138.5 mAh g$^{-1}$ for complete Na removal from Na$_2$Ni$_{2-x}$Co$_x$TeO$_6$. The operating temperature (25–75 °C) was controlled by EXPEC thermostat chambers (SU–242 or ST-110).

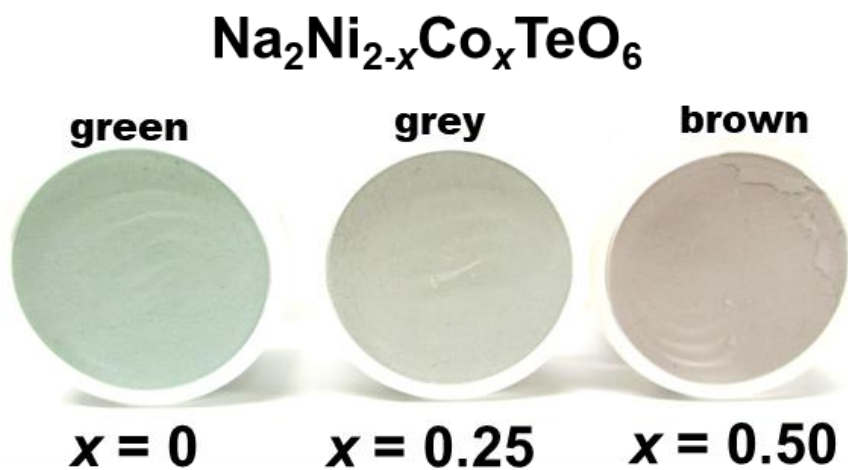

**Figure S1**. Variation in the color of Na$_2$Ni$_{2-x}$Co$_x$TeO$_6$ ($x$ = 0, 0.25 and 0.50) honeycomb layered oxides, indicating color change with partial substitution of cobalt in Na$_2$Ni$_2$TeO$_6$.

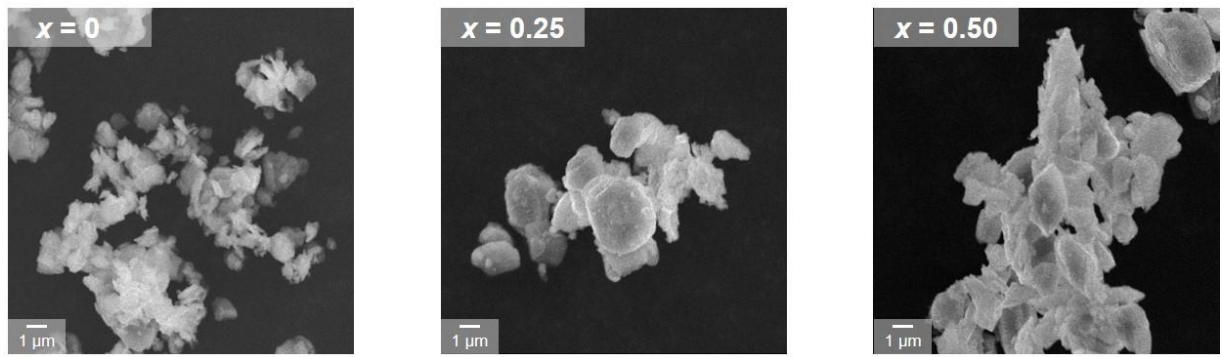

**Figure S2**. Scanning electron microscopy (SEM) picture of the as-synthesized Na$_2$Ni$_{2-x}$Co$_x$TeO$_6$ ($x$ = 0, 0.25, and 0.50) indicating a wide micrometric size particle distribution.

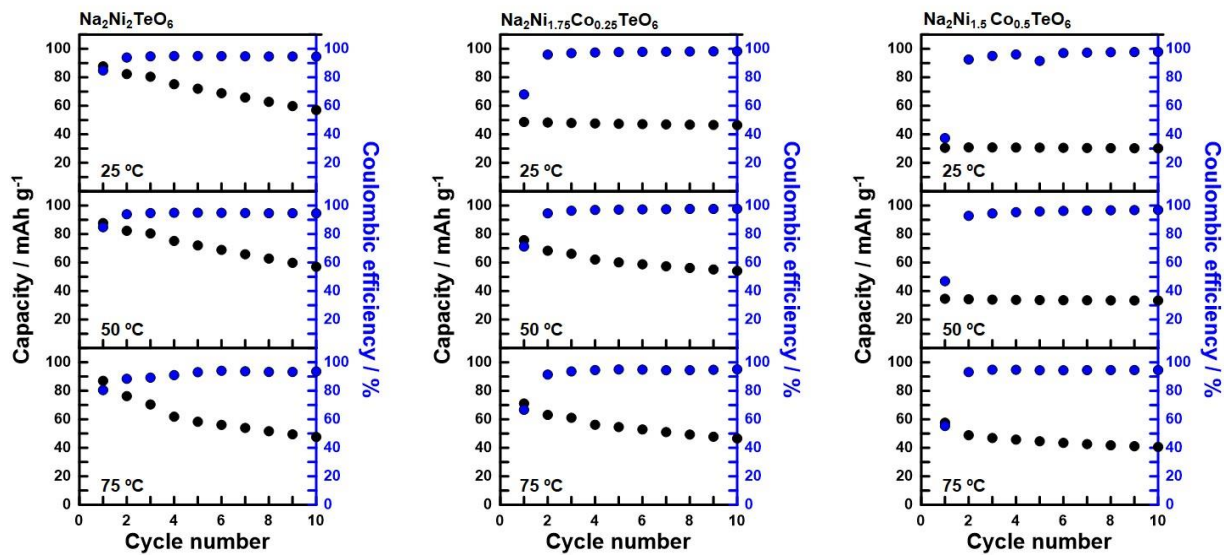

**Figure S3**. Capacity retention and coulombic efficiency of $Na_2Ni_{2-x}Co_xTeO_6$ ($x$ = 0, 0.25, and 0.50) during galvanostatic (dis)charging at a current density equivalent to 6.925 mA g$^{-1}$ (C/20). The measurements were performed in the voltage range of 2.5–4.35 V at 25, 50, and 75 °C.

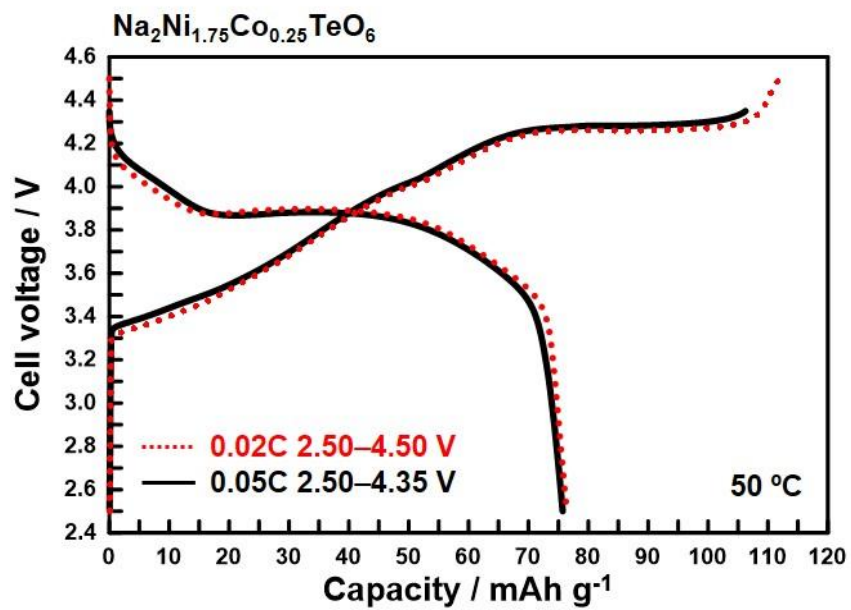

**Figure S4**. Galvanostatic charge–discharge curves of Na$_2$Ni$_{1.75}$Co$_{0.25}$TeO$_6$ measured at 50 °C in the voltage ranges of (i) 2.5–4.35 V at a current density of (i) C/20 (6.925 mA g$^{-1}$) and (ii) at a higher cut-off voltage of 2.5–4.5 V at C/50 (2.77 mA g$^{-1}$). Only the first cycle is shown for clarity.

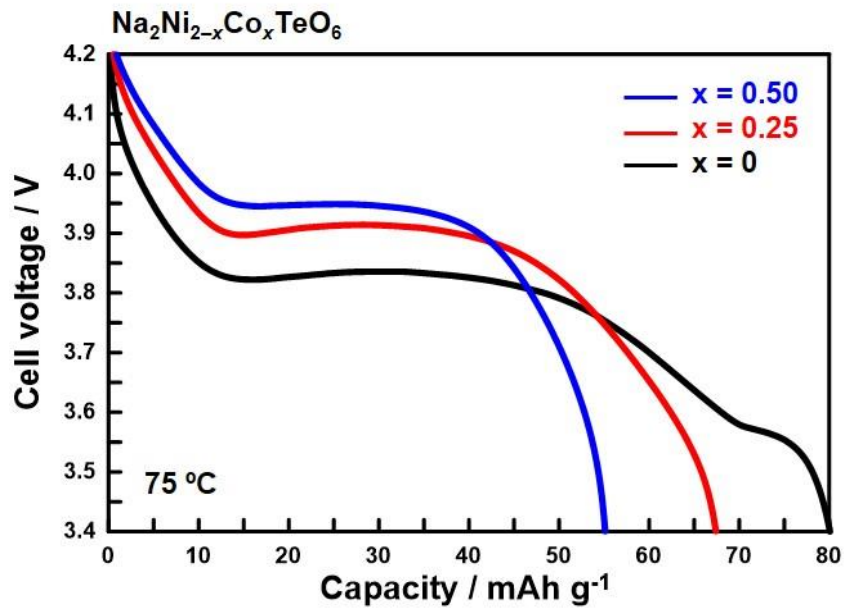

**Figure S5**. Comparison of the discharge capacity curves of Na$_2$Ni$_{2-x}$Co$_x$TeO$_6$ ($x$ = 0, 0.25, and 0.50) at a current density equivalent to C/20 (6.925 mA g$^{-1}$) at 75 °C

**Table S1**. Atomic coordinates ($x$, $y$, $z$), Wyckoff symbols and occupancies ($g$) obtained from Rietveld refinement of high-resolution conventional X-ray diffraction data for as-prepared $Na_2Ni_{1.5}Co_{0.5}TeO_6$ ($x=0.50$ in $Na_2Ni_{2-x}Co_xTeO_6$) indexed in the centrosymmetric space group $P6_3/mcm$ (hexagonal) with lattice constants $a = 5.2183(1)$ Å, $c = 11.1449(3)$ Å, and $V = 262.8(1)$ Å$^3$.

| Atom | Wyckoff | $x$ | $y$ | $z$ | $g$ |
|---|---|---|---|---|---|
| Te | 2b | 0 | 0 | 0 | 1 |
| Co | 4d | 2/3 | 1/3 | 0 | 0.25 |
| Ni | 4d | 2/3 | 1/3 | 0 | 0.75 |
| O | 12k | 0.6445(10) | 0.6445(10) | 0.5635(5) | 1 |
| Na1 | 6g | 0.3565(42) | 0 | 1/4 | 0.403 |
| Na2 | 4c | 1/3 | 2/3 | 1/4 | 0.184 |
| Na3 | 2a | 0 | 0 | 1/4 | 0.029 |

$R_{wp}=9.57\%$  $R_p=6.54\%$  $\chi^2=2.24$

* Preferred orientation ($hkl$) : ($00l$)

**Table S2**. Atomic coordinates ($x$, $y$, $z$), Wyckoff symbols and occupancies ($g$) obtained by Rietveld refinement of high-resolution conventional X-ray diffraction data for as-prepared $Na_2Ni_{1.75}Co_{0.25}TeO_6$ ($x$=0.25 in $Na_2Ni_{2-x}Co_xTeO_6$) indexed in the centrosymmetric space group $P6_3/mcm$ (hexagonal) with lattice constants $a$ = 5.2151(1) Å, $c$ = 11.1514(4) Å, and $V$ = 262.6(1) Å$^3$.

| Atom | Wyckoff | $x$ | $y$ | $z$ | $g$ |
| --- | --- | --- | --- | --- | --- |
| Te | 2b | 0 | 0 | 0 | 1 |
| Co | 4d | 2/3 | 1/3 | 0 | 0.125 |
| Ni | 4d | 2/3 | 1/3 | 0 | 0.875 |
| O | 12k | 0.6517(7) | 0.6517(7) | 0.5813(3) | 1 |
| Na1 | 6g | 0.3394(25) | 0 | 1/4 | 0.401(2) |
| Na2 | 4c | 1/3 | 2/3 | 1/4 | 0.177(5) |
| Na3 | 2a | 0 | 0 | 1/4 | 0.161(3) |

$R_{wp}$=8.51%   $R_p$=6.22%   $\chi^2$=2.03

* Preferred orientation ($hkl$) : (00$l$)

**Table S3**. Atomic coordinates ($x$, $y$, $z$), Wyckoff symbols and occupancies ($g$) from Rietveld refinement of high-resolution conventional X-ray diffraction data for as-prepared $Na_2Ni_2TeO_6$ ($x=0$ in $Na_2Ni_{2-x}Co_xTeO_6$) indexed in the centrosymmetric space group $P6_3/mcm$ (hexagonal) with lattice constants $a$ = 5.2049(1) Å, $c$ = 11.1505(5) Å, and $V$ = 261.6(1) Å$^3$.

| Atom | Wyckoff | $x$ | $y$ | $z$ | $g$ |
| --- | --- | --- | --- | --- | --- |
| Te | 2b | 0 | 0 | 0 | 1 |
| Ni | 4d | 2/3 | 1/3 | 0 | 1 |
| O | 12k | 0.6509(7) | 0.6509(7) | 0.5825(3) | 1 |
| Na1 | 6g | 0.3431(24) | 0 | 1/4 | 0.421(2) |
| Na2 | 4c | 1/3 | 2/3 | 1/4 | 0.179(3) |
| Na3 | 2a | 0 | 0 | 1/4 | 0.262 |

$R_{wp}$=8.12%   $R_p$=6.00%   $\chi^2$=1.95